# Reformulating hyperdynamics without a transition state theory dividing surface

Woo Kyun Kim

*Department of Materials Science and Engineering, The Johns Hopkins University, Baltimore, Maryland 21218*

Michael L. Falk

*Department of Materials Science and Engineering, The Johns Hopkins University, Baltimore, Maryland 21218*

*Department of Mechanical Engineering, The Johns Hopkins University, Baltimore, Maryland 21218*

*Department of Physics and Astronomy, The Johns Hopkins University, Baltimore, Maryland 21218*

Reformulating hyperdynamics without using a transition state theory (TST) dividing surface makes it possible to accelerate conventional molecular dynamics (MD) simulation using a broader range of bias potentials. A new scheme to calculate the boost factor is also introduced that makes the hyperdynamics method more accurate and efficient. Novel bias potentials using the hyper-distance and the potential energy slope and curvature along the direction vector from a minimum to a current position can significantly reduce the computational overhead required. Results simulating an atomic force microscope (AFM) system validate the new methodology.

## I. INTRODUCTION

When a dynamical system is confined in a potential energy basin separated by large energy barriers (>> $k_BT$), the system stays in this basin for a very long time compared to a typical atomic vibrational period before hopping to other basins. If the potential energy has multiple such minima (metastable states), the system evolves through infrequent transitions from one metastable state to another. In physical systems dominated by infrequent events, information about the waiting times at each state, the transition mechanisms leading to other states, and their relative probabilities is essential to understand and predict their behavior.

Direct dynamics simulation methods like molecular dynamics (MD) [1] do not assume any prior knowledge about how a system will evolve in time and can therefore be used to simulate as yet unknown transition mechanisms. However, since the overall time scale accessible by the conventional MD simulation method is restricted by the atomic vibrational period due to issues of numerical stability, it has been difficult to simulate these infrequent event problems using the MD methodology. Alternatively, if we were able to enumerate every transition mechanism and its rate for all the states the system visits during its evolution, this would allow us to advance the system from one state to another without following detailed trajectories in configuration space. This is the fundamental assumption of Kinetic Monte Carlo (KMC).[2] Moreover, when two adjacent minima are known, the transition rate between these two states can be calculated using transition state theory (TST) [3-8] if a proper dividing surface that the system crosses in transitions can be constructed. However, determining all transition mechanisms becomes more and more intractable as system complexity increases. Therefore, performing dynamics simulations on time scales reachable by KMC has long been a goal.

In recent years several novel methods to extend the MD time scale have been proposed. These methods include hyperdynamics,[9, 10] the parallel replica method,[11] and temperature-accelerated dynamics (TAD).[12] In the parallel replica method multiple replicas of a given system are simultaneously simulated on different processors so that the transitions can be accelerated up to a factor corresponding to the number of processors.[11] TAD uses a higher temperature MD simulation to more effectively detect transition pathways at a given lower temperature, but it assumes that the harmonic TST [13] holds and requires the saddle point to be found for each transition pathway.[12] A more detailed review of these methods is also found in Ref. 14.

Taking into account both the achievable boost factor and the degree of approximation, hyperdynamics is perhaps the most attractive acceleration method. In hyperdynamics, [9] a given potential energy function is modified such that the energy barriers are reduced while the characteristic dynamics are preserved. In principle hyperdynamics simulation can advance the system at an accelerated pace while preserving the correct relative transition probabilities under the assumption that the TST transition rates are equivalent to the actual rates. Furthermore, the acceleration rate can be calculated concurrently during the simulation. However, constructing a computationally efficient bias potential for a hyperdynamics simulation can be difficult. In Voter's original bias potentials the lowest eigenvalue and the corresponding eigenvector of the Hessian matrix are used, [9, 10] but calculating the eigenvalue and its derivative require significant computational overhead. To avoid these excessive computations several simplified bias potentials have been proposed,[14] and recently Miron and Fichthorn proposed a bias potential called the bond-boost method using the bond length changes to detect a transition without significant computational overhead.[15] The bond-boost method utilizes the characteristic of bond-breaking that most solid-state systems undergo when making transitions to construct a bias potential. However, this method gives rise to a significant force discontinuity due to the envelope function introduced to enforce that the bias potential is zero at the dividing surface. It can also introduce spurious energetic minima within an individual bond.

Although several bias potentials have been proposed thus far, all either have significant computational overhead, which degrades the achieved boost factor, or lack the generality required to detect a diverse range of transitions. Thus, finding a bias potential that provides a large boost without significant overhead remains a challenging problem.

In this manuscript we begin by reviewing the fundamentals of the hyperdynamics method and then proceed to reformulate the method in a rigorous way that obviates the need to construct a TST dividing surface. Special emphasis is put on the importance of accurately calculating the boost factor. This will set the stage for devising new bias potentials using local variables in place of or in addition to the lowest eigenvalue of the Hessian matrix. Finally, the new methodology is tested with an atomic force micro scope (AFM) system modeled using the Lennard-Jones potential.

## II. REVIEW

### A. Rare events and transition rate

Let us consider a system in an ensemble with constant boundary conditions such as the canonical ensemble (NVT). The characteristics of this system can be described by the Hamiltonian $H$ defined by

$$H(\vec{r},\vec{v}) = V(\vec{r}) + K(\vec{v}) \ , \tag{1}$$

where $\vec{r}$ is the 3$N$-dimensional position vector in the configuration space ($N$ is the total number of particles), $\vec{v}$ is the 3$N$-dimensional velocity vector, $V$ is the potential energy, and $K$ is the kinetic energy. The probability density distribution $\rho(\vec{r},\vec{v})$ in the phase space is given by

$$\rho(\vec{r},\vec{v}) = \frac{1}{Z}\exp(-\beta H(\vec{r},\vec{v})) \ , \tag{2}$$

where $Z$ is the partition function defined by $\int d\,\vec{r}\int d\,\vec{v}\ e^{-\beta H}$ and $\beta = 1/k_B T$ ; $k_B$ is the Boltzmann constant and $T$ is the temperature. Then, the ensemble average of any observable $O(\vec{r},\vec{v})$ is expressed as

$$\langle O(\vec{r},\vec{v})\rangle = \int d\,\vec{r}\int d\,\vec{v}\ O(\vec{r},\vec{v})\rho(\vec{r},\vec{v}) = \frac{1}{Z}\int d\,\vec{r}\int d\,\vec{v}\ O(\vec{r},\vec{v})e^{-\beta H} \quad . \tag{3}$$

The dynamics of this system $\{\vec{r}(t),\vec{v}(t)\}$, a trajectory in the phase space satisfying the probability density in Eq. (2), can be modeled, for example, by coupling an isolated system to the Nosé-Hoover thermostat.[16-18]

If the potential energy has multiple local minima separated by high energy barriers (>> $k_BT$), each local minimum in configuration space and its neighborhood, around which the probabilities are concentrated, defines a metastable state and the characteristic dynamics of the system consists of infrequent transitions from one metastable state to another as illustrated in Fig. 1. If the waiting time at each state is comparable to the observation time, the time average of an observable is not given by Eq. (3), but depends on the specific transition paths the system follows. We assume that the metastable states each have a corresponding non-overlapping configuration space volume, called a state volume, so that they are distinguishable from each other and do not share any position vector as illustrated in Fig. 1.

Now we shall describe how to calculate the waiting times and the relative transition probabilities. Let us consider a transition from a state A to one of two neighboring states B and C (hereafter we use capital letters to refer to either a state or a state volume unless an ambiguity would arise). Then, a transition from state A occurs only when the system trajectory enters the state volume B or C after leaving the state volume A rather than re-entering itself. The system can cross and recross the boundary of the state volume A several times before making a transition (see Fig. 1). Thus, when we say that the system enters, leaves, and stays at state A, we refer to the initial entrance event to the state volume A, the final leaving event, and the time period between these two events, respectively. This definition of a transition is essentially the same as that in Ref. 8.

The waiting time $t_A$ is defined as the total period of time since the trajectory initially enters the state volume A until the trajectory finally leaves the state volume A. If the system leaves the state volume A only when making a transition, the waiting time is the same as the time period the system spends inside of A. However, because in practice it is difficult to construct such an ideal state volume, we consider the possible recrossings of the boundary of A before the trajectory finally leaves it. Then, $t_A$ can be expressed as

$$t_A = t_{A,i} + t_{A,o} \quad , \tag{4}$$

where $t_{A,i}$ is the total time the system spends inside of A and $t_{A,o}$ is the total time the system spends outside of A before making a transition. The totality of the positions outside of A where the system can visit before making a transition is referred to as the transition region of A and is denoted as $\widetilde{\mathrm{A}}$. Note that if the state volume A is well chosen $t_{A,i} >> t_{A,o}$.

Since in infrequent events the waiting times are long and uncorrelated we can assume that $t_A$ has a Poisson distribution expressed as [8]

$$\mathrm{P}\,(t_A) = R_A \exp\,(-R_A\, t_A) \ , \tag{5}$$

where $\mathrm{P}(\cdots)$ is a probability density function, and $R_A$ is a rate constant that characterizes the transition. We can also show that the following relation holds.

$$\mathrm{E}(t_A) = \int_0^\infty t\ \mathrm{P}(t)\, dt = 1/R_A \ , \tag{6}$$

where $\mathrm{E}(\cdots)$ is an expectation value. If we can construct an infinite trajectory such that the system visits and escapes from state A a large number of times, the average waiting time can be calculated by

$$\bar{t}_A \equiv \lim_{\tau\to\infty} \frac{\sum_{k=1}^{N_{A\to}} t_A^k}{N_{A\to}} = \mathrm{E}(t_A) \quad , \tag{7}$$

where $\tau$ is the total time elapsed by the system; $N_{A\to}$ is the total number of the transitions from state A in this time interval; $t_A^k$ is the waiting time during the $k$th stay at state A. Hereafter we use bars to refer to the average of samples obtained in this infinite trajectory as in Eq. (7). By rearranging Eq. (7), we obtain

$$\bar{t}_A = \lim_{\tau\to\infty} \frac{\sum_{k=1}^{N_{A\to}} t_A^k}{\tau} \frac{\tau}{N_{A\to}} = \frac{p_A}{\upsilon_{A\to}} \quad , \tag{8}$$

where

$$\upsilon_{A\to} \equiv \lim_{\tau\to\infty} \frac{N_{A\to}}{\tau} \quad , \tag{9}$$

$$p_A \equiv \lim_{\tau\to\infty} \frac{\sum_{k=1}^{N_{A\to}} t_A^k}{\tau} \quad . \tag{10}$$

where $\upsilon_{A\to}$ is the mean frequency of the transition from state A and $p_A$ is the ratio of the total waiting times at state A to the entire time interval. $p_A$ can be expressed as

$$p_A = \lim_{\tau\to\infty} \frac{\sum_{k=1}^{N_{A\to}} t_A^k}{\tau} = \lim_{\tau\to\infty} \frac{\sum_{k=1}^{N_{A\to}} t_{A,i}^k + \sum_{k=1}^{N_{A\to}} t_{A,o}^k}{\tau} = p_{A,i} + p_{A,o} \quad , \tag{11}$$

where

$$p_{A,i} \equiv \lim_{\tau\to\infty} \frac{\sum_{k=1}^{N_{A\to}} t_{A,i}^k}{\tau} \quad \text{and} \quad p_{A,o} \equiv \lim_{\tau\to\infty} \frac{\sum_{k=1}^{N_{A\to}} t_{A,o}^k}{\tau} \quad . \tag{12}$$

By definition, the state volume A does not overlap with other state volumes so that in an ergodic system we have the following relation,

$$p_{A,i} = \frac{\int_A d\vec{r}\, e^{-\beta V}}{\int d\vec{r}\, e^{-\beta V}} \quad . \tag{13}$$

However, in general $\tilde{\mathrm{A}}$ can overlap with $\tilde{\mathrm{B}}$ or $\tilde{\mathrm{C}}$ so that we have

$$p_{A,o} = \lim_{\tau\to\infty} \frac{\sum_{k=1}^{N_{A\to}} t_{A,o}^{k}}{\tau} \le \frac{\int_{\tilde{A}} d\vec{r}\ e^{-\beta V}}{\int d\vec{r}\ e^{-\beta V}} \quad \text{or} \quad p_{A,o} = c_A \times \frac{\int_{\tilde{A}} d\vec{r}\ e^{-\beta V}}{\int d\vec{r}\ e^{-\beta V}} \quad (c_A \le 1) \ . \tag{14}$$

If $\widetilde{\mathrm{A}}$ does not overlap with other transition regions or the probabilities of the overlapping regions are negligible, then $c_A \approx 1$. In the derivations below we ignore the possibility that $c_A \ne 1$. The procedures to calculate the mean frequency of transition is discussed in the next section.

Now we consider the relative transition probabilities for $\mathrm{A} \to \mathrm{B}$ and $\mathrm{A} \to \mathrm{C}$. We can define the mean frequencies for these specific transitions as

$$\upsilon_{A\to B} = \lim_{\tau\to\infty} \frac{N_{A\to B}}{\tau} \quad \text{and} \quad \upsilon_{A\to C} = \lim_{\tau\to\infty} \frac{N_{A\to C}}{\tau} \quad , \tag{15}$$

where $N_{A\to B}$ and $N_{A\to C}$ are the total number of transitions from state A to state B and state C in the time interval [0, $\tau$] respectively. Note that since $N_{A\to} = N_{A\to B} + N_{A\to C}$, $\upsilon_{A\to} = \upsilon_{A\to B} + \upsilon_{A\to C}$. Then, the relative transition probabilities are given by

$$\frac{p_{A\to C}}{p_{A\to B}} = \lim_{\tau\to\infty} \left( \frac{N_{A\to C}}{N_{A\to B}} \right) = \frac{\upsilon_{A\to C}}{\upsilon_{A\to B}} \quad , \tag{16}$$

where $p_{A\to B}$ and $p_{A\to C}$ refers to the relative probabilities to the transition $\mathrm{A} \to \mathrm{B}$ and $\mathrm{A} \to \mathrm{C}$ such that

$$p_{A\to B} + p_{A\to C} = 1 \ . \tag{17}$$

Therefore, all the information needed to understand a dynamical system undergoing infrequent events can be obtained from the mean transition frequencies.

### B. Transition state theory and dynamical corrections

In principle, to obtain the mean frequencies of transition in Eq. (9) and (15), we need to observe a statistical number of transitions from the state in question so that we can count the total number of transitions to each neighbor and measure the waiting time. Practically it is hard to observe even a single transition if we use a conventional dynamics scheme. When two states, usually referred as the reactant state and product state, are known, transition state theory can be used to calculate the mean transition frequency between these two states. In this section we briefly review TST transition rates and the dynamical corrections to these rates.

Given two metastable states A and B, we calculate the mean frequency $\upsilon_{A\to B}$ of the transition from A to B. First, we construct a dividing surface *S* between the corresponding state volumes, which is neither the boundary of A ($\partial A$) nor B ($\partial B$) (see Fig. 1), then the mean frequency $\upsilon_S$ of crossing *S* from A to B is given by [8]

$$\upsilon_S = \frac{1}{2}\frac{\int_S dS \int d\vec{v}\,|v_n|\,e^{-\beta(V+K)}}{Z} \quad , \tag{18}$$

where $v_n$ is the velocity normal to the dividing surface *S* and the prefactor 1/2 is introduced to account for the transition A$\rightarrow$B only. Note that $\upsilon_S \geq \upsilon_{A\rightarrow B}$ because the system can recross the dividing surface several times before making a transition. No transition has occurred if the trajectory leaving $\partial A$ re-enters A without visiting B. Although we can find the optimized dividing surface which minimizes $\upsilon_S$ (variational TST [4]), there is still a chance of such an event.

To account for the possible recrossing of *S* the dynamical correction factor $f_d$ has been introduced, [5, 6, 8, 19] which is defined as

$$f_d = \frac{\upsilon_{A\rightarrow B}}{\upsilon_S} \quad . \tag{19}$$

Although we cannot directly calculate $\upsilon_{A\rightarrow B}$, we can calculate $f_d$ as follows. [8] First, we sample a large number of points on the dividing surface *S* and initialize velocities according to the correct phase space distribution. We initiate a pair of trajectories starting from the same point on the dividing surface, one of which has the chosen outward velocity and the other of which has the opposite inward velocity. Then, the ratio of the number of trajectories whose outward pair enters B and whose inward pair enters A without returning to the dividing surface to the total number of trajectories gives the approximate dynamical correction factor.

### C. Hyperdynamics

Although the original hyperdynamics formulation is based on the TST transition rate, [9] the actual transition rate or the mean frequency of transition does not depend on the choice of the dividing surface. In the previous section a dividing surface is used to calculate $\upsilon_S$ and $f_d$, but this is merely a construction, and the product of these must not depend on the choice of a dividing surface. As we will show it is possible to formulate the hyperdynamics method without using a TST dividing surface. In the hyperdynamics method it is not necessary to directly calculate the transition rate. Instead, we need only to be able to accurately compute the ratio of the transition rates in two different potentials and this can be done without resorting to the construction of a dividing surface.

First, by rearranging Eq. (9) we have

$$\upsilon_{A\to} = \lim_{\tau\to\infty} \frac{N_{A\to}}{\tau} = \lim_{\tau\to\infty} \frac{N_{\partial A\to}}{\tau} \frac{N_{A\to}}{N_{\partial A\to}} = \upsilon_{\partial A\to} \times f_{A\to} \quad , \tag{20}$$

where

$$\upsilon_{\partial A\to} = \lim_{\tau\to\infty} \frac{N_{\partial A\to}}{\tau} \quad , \tag{21}$$

$$f_{A\to} = \lim_{\tau\to\infty} \frac{N_{A\to}}{N_{\partial A\to}} \quad , \tag{22}$$

where $N_{\partial A\to}$ is the total number of trajectories crossing $\partial A$ in the time interval [0, $\tau$]. $\upsilon_{\partial A\to}$ is expressed as

$$\upsilon_{\partial A\to} = \frac{1}{2} \frac{\int_{\partial A} dS \int d\vec{v}\, |v_n|\, e^{-\beta(V+K)}}{Z} \quad . \tag{23}$$

Using $N_{A\to B}$ and $N_{A\to C}$, $f_{A\to}$ can be expressed as

$$f_{A\to} = f_{A\to B} + f_{A\to C} \tag{24}$$

where

$$f_{A\to B} = \lim_{\tau\to\infty} \frac{N_{A\to B}}{N_{\partial A\to}} \quad \text{and} \quad f_{A\to C} = \lim_{\tau\to\infty} \frac{N_{A\to C}}{N_{\partial A\to}} \quad . \tag{25}$$

Then, we have

$$\frac{p_{A\to C}}{p_{A\to B}} = \lim_{\tau\to\infty} \left( \frac{N_{A\to C}}{N_{A\to B}} \right) = \frac{f_{A\to C}}{f_{A\to B}} \quad . \tag{26}$$

Note that $f_{A\to B}$ and $f_{A\to C}$ depend only on the potential energy values at $\partial A$ and in the transition region, which is outside of A. To prove this, imagine the distribution of crossing points of the trajectories from inward with $\partial A$. Since these points also belong to the phase space, they must satisfy the following distribution

$$\rho_S(\vec{r}) = \frac{e^{-\beta V}}{Z_S} \quad , \tag{27}$$

where $Z_S = \int_S e^{-\beta V}\, dS$. Moreover, whether a specific trajectory will return to $\partial A$ or enter other states after it leaves $\partial A$ is completely determined by the potential energy values in transition regions. Thus, if two potentials have the same values in these regions ($\partial A$ and the transition region), $f_{A\to B}$ and $f_{A\to C}$ remain unchanged and the relative transition probabilities in these two potentials are identical. Note that although with these two potentials $N_{\partial A\to}$ changes, $f_{A\to B}$, the ratio of $N_{A\to B}$ to $N_{\partial A\to}$, remains unchanged. This is the fundamental basis of the hyperdynamics method.

Now we assume that for a given potential *V*, we can construct a biased potential $V_b$ such that

$$V(\vec{r}) = V_b(\vec{r}) + \Delta V \quad , \tag{28}$$

where a bias potential $\Delta V$ satisfies the following condition

$$\Delta V(\vec{r}) = \begin{cases} > 0\,, & in\ A \\ = 0\,, & along\ \partial A\ and\ outside\ of\ A \end{cases} \quad . \tag{29}$$

Then, the ratio of the transition rates in *V* and $V_b$, the boost factor $\alpha$, is given by

$$\alpha = \frac{(R_A)_b}{R_A} = \frac{\overline{t_A}}{\overline{(t_A)_b}} = \frac{p_A / \upsilon_{A\to}}{(p_A)_b / (\upsilon_{A\to})_b} \qquad \text{from Eq. (8)}$$

$$= \frac{p_A / \upsilon_{\partial A\to} f_{A\to}}{(p_A)_b / (\upsilon_{\partial A\to})_b (f_{A\to})_b} \qquad \text{from Eq. (20)}$$

$$= \frac{p_A / \upsilon_{\partial A\to}}{(p_A)_b / (\upsilon_{\partial A\to})_b} \qquad \text{from } f_{A\to} = (f_{A\to})_b$$

$$= \frac{\int_{A+\tilde{A}} d\vec{r}\ e^{-\beta V}}{\int_{A+\tilde{A}} d\vec{r}\ e^{-\beta V_b}} \qquad \text{from Eq. (13), (14) and (23).} \tag{30}$$

Hereafter the subscript *b* means that the property is obtained in the biased potential.

It is apparent that the average waiting time in the original potential can be recovered from the average waiting time in the biased potential as shown in Eq. (30). Now we consider the probability density function of the waiting time. We define the recovered time $\eta$ as $\alpha \times (t_A)_b$. The probability density function of $\eta$ is given by

$$\mathrm{P}(\eta) = \frac{d\,(t_A)_b}{d\eta} \mathrm{P}((t_A)_b)$$

$$= \frac{1}{\alpha} \mathrm{P}((t_A)_b) \qquad \text{from } \frac{d\,(t_A)_b}{d\eta} = \frac{1}{\alpha}$$

$$= \frac{1}{\alpha} \times (R_A)_b \exp\big((R_A)_b \times (t_A)_b\big) \qquad \text{from Eq. (5) for } (t_A)_b$$

$$= R_A \exp(R_A\ \eta) \qquad \text{from Eq. (30) .} \tag{31}$$

Therefore, the stochastic outcome of the waiting time in the original potential can exactly be replaced by the recovered time $\eta\,(=\alpha\times(t_A)_b\,)$.

The state volume A can be well-defined, but $\tilde{A}$ may not. Thus, we derive an alternative formula. First, we define the biased boost factor as

$$\alpha_b=\frac{\overline{t_{A,i}}}{\overline{(t_{A,i})_b}}\quad . \tag{32}$$

Then, using the same derivation in Eq. (30) we have

$$\alpha_b=\frac{\int_A d\vec{r}\;e^{-\beta V}}{\int_A d\vec{r}\;e^{-\beta V_b}}\quad . \tag{33}$$

We can also show that $\overline{t_{A,o}}=\overline{(t_{A,o})_b}$. Then, the boost factor becomes

$$\alpha=\frac{\overline{t_A}}{\overline{(t_A)_b}}=\frac{\overline{t_{A,i}}+\overline{t_{A,o}}}{\overline{(t_{A,i})_b}+\overline{(t_{A,o})_b}}=\frac{\alpha_b\times\overline{(t_{A,i})_b}+\overline{(t_{A,o})_b}}{\overline{(t_{A,i})_b}+\overline{(t_{A,o})_b}}$$
$$=(p_{A,i})_b\times\alpha_b+(p_{A,o})_b\;, \tag{34}$$

where

$$(p_{A,i})_b=\frac{\overline{(t_{A,i})_b}}{\overline{(t_{A,i})_b}+\overline{(t_{A,o})_b}}=\frac{\int_A d\vec{r}\;e^{-\beta V_b}}{\int_{A+\tilde{A}} d\vec{r}\;e^{-\beta V_b}}\quad , \tag{35}$$

$$(p_{A,o})_b=\frac{\overline{(t_{A,o})_b}}{\overline{(t_{A,i})_b}+\overline{(t_{A,o})_b}}=\frac{\int_{\tilde{A}} d\vec{r}\;e^{-\beta V_b}}{\int_{A+\tilde{A}} d\vec{r}\;e^{-\beta V_b}}\quad . \tag{36}$$

In the actual hyperdynamics simulations $(p_{A,i})_b$ and $(p_{A,o})_b$ can be approximated by $(p_{A,i})_b^*$ and $(p_{A,o})_b^*$ as calculated from the time of residence inside and outside state A in the boosted simulation, i.e.

$$(p_{A,i})_b^*\approx\frac{(t_{A,i})_b}{(t_{A,i})_b+(t_{A,o})_b}\quad\text{and}\quad(p_{A,o})_b^*\approx\frac{(t_{A,o})_b}{(t_{A,i})_b+(t_{A,o})_b}\quad . \tag{37}$$

Then, we can show that $\overline{t_A}=\overline{\alpha\times(t_A)_b}=\overline{\alpha_b\times(t_{A,i})_b+(t_{A,o})_b}$. Thus, without knowing the actual boost factor $\alpha$ we can recover the original waiting time with the biased boost factor $\alpha_b$.

In the same way in Eq. (34) we can also show that

$$\alpha=\frac{1}{p_{A,i}/\alpha_b+p_{A,o}}<\frac{1}{p_{A,o}}\quad . \tag{38}$$

In the case that $p_{A,i}$, $p_{A,o}$, and $\alpha$ are known, $(p_{A,i})_b$ and $(p_{A,o})_b$ can be predicted using the following relations.

$$(p_{A,i})_b = \frac{\int_{A+\tilde{A}} d\vec{r}\, e^{-\beta V}}{\int_{A+\tilde{A}} d\vec{r}\, e^{-\beta V_b}} \frac{\int_{A} d\vec{r}\, e^{-\beta V_b}}{\int_{A} d\vec{r}\, e^{-\beta V}} \frac{\int_{A} d\vec{r}\, e^{-\beta V}}{\int_{A+\tilde{A}} d\vec{r}\, e^{-\beta V}} = \frac{\alpha}{\alpha_b} \times p_{A,i} \quad , \tag{39}$$

$$(p_{A,o})_b = \frac{\int_{A+\tilde{A}} d\vec{r}\, e^{-\beta V}}{\int_{A+\tilde{A}} d\vec{r}\, e^{-\beta V_b}} \frac{\int_{\tilde{A}} d\vec{r}\, e^{-\beta V_b}}{\int_{\tilde{A}} d\vec{r}\, e^{-\beta V}} \frac{\int_{\tilde{A}} d\vec{r}\, e^{-\beta V}}{\int_{A+\tilde{A}} d\vec{r}\, e^{-\beta V}} = \alpha \times p_{A,o} \quad . \tag{40}$$

Note that although $p_{A,i} >> p_{A,o}$, with aggressive boosting $(p_{A,o})_b$ can be comparable to $(p_{A,i})_b$.

Finally, the ensemble averages of an observable $O(\vec{r})$ in the original potential and the biased potential have the following relation

$$\left\langle O(\vec{r}) \right\rangle = \frac{1}{\alpha} \left\langle O(\vec{r})\, e^{+\beta \Delta V} \right\rangle_b \quad , \tag{41}$$

where

$$\left\langle f(\vec{r}) \right\rangle = \frac{\int_{A+\tilde{A}} f(\vec{r}) e^{-\beta V} d\vec{r}}{\int_{A+\tilde{A}} e^{-\beta V} d\vec{r}} \quad , \tag{42}$$

$$\left\langle f(\vec{r}) \right\rangle_b = \frac{\int_{A+\tilde{A}} f(\vec{r}) e^{-\beta V_b} d\vec{r}}{\int_{A+\tilde{A}} e^{-\beta V_b} d\vec{r}} \quad . \tag{43}$$

Therefore, the ensemble average of an observable in the original potential can be obtained from the ensemble average in the biased potential using Eq. (41) without performing a simulation with the original potential.

## III. CRITICAL ISSUES OF HYPERDYNAMICS

### A. Boost factor

According to Voter's original prescription [9], the time $t$ in the original potential can be recovered from the time $t_b$ elapsed in the potential $V_b$ modified with a bias potential $\Delta V$ using

$$t = \sum_{i=1}^{N_{TOT}} \Delta t_{MD}\, e^{+\beta\, \Delta V[\vec{r}(t_i)]} \quad , \tag{44}$$

where $N_{TOT}$ is the total number of MD steps within one transition event, $\Delta t_{MD}$ is the time interval for numerical integration, and $t_i$ is the time at *i*th MD step ($= i \times \Delta t_{MD}$). We can prove Eq. (44) using the recovered time ($= t_b \times \alpha$ ). By rearranging Eq. (30) we obtain the following relation

$$\alpha = \frac{\int_{A+\widetilde{A}} e^{-\beta V} \, d\vec{r}}{\int_{A+\widetilde{A}} e^{-\beta V_b} \, d\vec{r}} = \frac{\int_{A+\widetilde{A}} e^{+\beta \Delta V} e^{-\beta V_b} \, d\vec{r}}{\int_{A+\widetilde{A}} e^{-\beta V_b} \, d\vec{r}} = \left\langle e^{+\beta \Delta V} \right\rangle_b \quad . \tag{45}$$

From Eq. (45), the boost factor can be approximated by

$$\alpha = \left\langle e^{+\beta \Delta V} \right\rangle_b \approx \frac{\sum_{i=1}^{N_{TOT}} e^{+\beta \Delta V[\vec{r}(t_i)]}}{N_{TOT}} \quad . \tag{46}$$

Note that the MD simulation should be performed with the biased potential, i.e., the force vector should be obtained from the biased potential, in order to calculate $\alpha$ using Eq. (46). Since the recovered time can be regarded as the original waiting time, we have

$$\begin{aligned} t &= \eta = t_b \times \alpha \\ &= N_{TOT} \times \Delta t_{MD} \times \alpha \qquad \text{from } \; t_b = N_{TOT} \times \Delta t_{MD} \\ &= \sum_{i=1}^{N_{TOT}} \Delta t_{MD} \, e^{+\beta \Delta V[\vec{r}(t_i)]} \qquad \text{from Eq. (46).} \end{aligned} \tag{47}$$

This agrees with the supposition in Eq. (44).

In the above derivation it is assumed that Eq. (46) holds, an assumption that largely depends on the total number of MD steps sampled in each escape event. If a bias potential is chosen too aggressively, the system will stay in the starting state for too short a time to obtain the boost factor $\alpha$ accurately. With an inaccurate boost factor, the recovered time loses its statistical meaning and its ability to approximate the original time. In an optimistic view expressed in Ref. 9, even with the aggressive choice, the accumulated time error after many transitions may vanish because the time error in each transition is not correlated with others. However, it is also likely that a bad choice of a bias potential can cause the time error in a biased way such that it always yields shorter or longer estimates at every transition. Moreover, using too conservative a bias potential is not desirable for obvious reasons.

Since we need the accurate boost factor for each transition, a better sampling scheme for the boost factor is needed and this sampling may be performed as a pre-simulation before the actual simulation. By rearranging Eq. (30), we have

$$\alpha = \frac{\int_{A+\tilde{A}} e^{-\beta V}\ d\vec{r}}{\int_{A+\tilde{A}} e^{-\beta V_b}\ d\vec{r}} = \frac{\int_{A+\tilde{A}} e^{-\beta W}\ d\vec{r}}{\int_{A+\tilde{A}} e^{-\beta V_b}\ d\vec{r}} \frac{\int_{A+\tilde{A}} e^{-\beta V}\ d\vec{r}}{\int_{A+\tilde{A}} e^{-\beta W}\ d\vec{r}}$$

$$= \frac{\int_{A+\tilde{A}} e^{-\beta W}\ d\vec{r}}{\int_{A+\tilde{A}} e^{-\beta (V_b - W)} e^{-\beta W}\ d\vec{r}} \frac{\int_{A+\tilde{A}} e^{+\beta (W-V)} e^{-\beta W}\ d\vec{r}}{\int_{A+\tilde{A}} e^{-\beta W}\ d\vec{r}}$$

$$= \frac{\left\langle e^{+\beta (W-V)} \right\rangle_W}{\left\langle e^{-\beta (V_b - W)} \right\rangle_W} \quad , \tag{48}$$

where $W$ is an intermediate bias potential. Thus, in principle we can use any type of intermediate bias potential (even the original potential) to calculate the boost factor.

Let us consider a case with a bias potential using a local variable $\lambda$ (e.g. the lowest eigenvalue of the Hessian matrix). From Eq. (48) we have the following relation

$$\alpha = \frac{1}{\left\langle e^{-\beta \Delta V} \right\rangle} = \frac{1}{\int_{-\infty}^{\infty} \rho(\lambda) e^{-\beta \Delta V(\lambda)} d\lambda} \quad , \tag{49}$$

where $\rho(\lambda)$ is the probability density function of $\lambda$ in $\mathrm{A} + \tilde{\mathrm{A}}$ with the original potential. If the state volume is defined by $\lambda < \lambda_{cr}$ (i.e. $\Delta V = 0$ when $\lambda > \lambda_{cr}$), then we have

$$\alpha = \frac{1}{\int_{-\infty}^{\infty} \rho(\lambda) e^{-\beta \Delta V(\lambda)} d\lambda} = \frac{1}{\int_{-\infty}^{\lambda_{cr}} \rho(\lambda) e^{-\beta \Delta V(\lambda)} d\lambda + \rho(\lambda > \lambda_{cr})} < \frac{1}{\rho(\lambda > \lambda_{cr})} \quad . \tag{50}$$

Therefore, the boost factor cannot exceed $1/\rho(\lambda > \lambda_{cr})$. If we use an intermediate bias potential $W = V + \Delta W(\lambda)$ using the same local variable $\lambda$ for the sampling, the probability density function $\rho_W(\lambda)$ in $W$ is given by

$$\rho_W(\lambda_O) = \frac{\int_{A+\tilde{A}} \delta(\lambda(\vec{r}) - \lambda_O) e^{-\beta W} d\vec{r}}{\int_{A+\tilde{A}} e^{-\beta W} d\vec{r}} = \frac{\int_{A+\tilde{A}} \delta(\lambda(\vec{r}) - \lambda_O) e^{-\beta (V + \Delta W(\lambda))} d\vec{r}}{\int_{A+\tilde{A}} e^{-\beta W} d\vec{r}}$$

$$= \frac{e^{-\beta \Delta W(\lambda_O)} \times \int_{A+\tilde{A}} \delta(\lambda(\vec{r}) - \lambda_O) e^{-\beta V} d\vec{r}}{\int_{A+\tilde{A}} e^{-\beta W} d\vec{r}}$$

$$= \frac{e^{-\beta \Delta W(\lambda_O)} \times \int_{A+\tilde{A}} \delta(\lambda(\vec{r}) - \lambda_O) e^{-\beta V} d\vec{r}}{\int_{A+\tilde{A}} e^{-\beta V} d\vec{r}} \frac{\int_{A+\tilde{A}} e^{-\beta V} d\vec{r}}{\int_{A+\tilde{A}} e^{-\beta W} d\vec{r}}$$

$$= \alpha_W \times e^{-\beta \Delta W(\lambda_O)} \times \rho(\lambda_O) \ , \tag{51}$$

where

$$\alpha_W = \frac{\int_{A+\tilde{A}} e^{-\beta V} d\vec{r}}{\int_{A+\tilde{A}} e^{-\beta W} d\vec{r}} \quad . \tag{52}$$

Due to the prefactor $e^{-\beta \Delta W}$ in Eq. (51), $\rho_W(\lambda)$ is biased such that the region in the configuration space where $\Delta W$ is larger is less sampled, and the region having smaller $\Delta W$ is more sampled. Thus, we can effectively sample the configuration space choosing a proper intermediate bias potential *W*. For some cases, more than two bias potentials can be used for sampling and the resultant probability density functions can be combined, for example, using the weighted histogram analysis method (WHAM) [20] as in umbrella sampling. Note that once we construct the probability density function in one specific potential, the boost factors in any bias potentials can be calculated. When using pre-simulations to calculate the boost factor, it is critical to have sufficient numbers of MD steps for sampling in order to obtain an accurate boost factor.

### B. State volumes and bias potentials

Since the original formulation of hyperdynamics [9] used the TST transition rate, the bias potential was constructed based on the TST dividing surface. The conventional TST dividing surface uses the steepest ascent/descent path described by

$$\frac{d\vec{r}}{ds} = \pm \frac{\nabla V}{|\nabla V|} \ , \tag{53}$$

where $ds \ (= |d\vec{r}|)$ is the arc length of the *3N*-dimensional curve in the configuration space. [21] All the points which can be led to a minimum by the steepest descent path comprise the state defined by the minimum, and the points on the boundary, which converge to one of the first-order saddle points instead of the minima, define the TST dividing surface. Instead of this conventional TST dividing surface Voter used an approximate dividing surface [9, 21] defined by

$$\vec{C}_1 \cdot \nabla V = 0 \quad \text{and} \quad \varepsilon_1 < 0 \ , \tag{54}$$

where $\varepsilon_1$ is the lowest eigenvalue of the Hessian matrix and $\vec{C}_1$ is the corresponding eigenvector. However, calculating the eigenvalue, the eigenvector, and their derivatives (to obtain force) is extremely time-consuming. As stated in the previous sections, our hyperdynamics formulation does not use a TST dividing surface, but depends on the boundaries of the state volumes. This gives more flexibility in how we define bias potentials. To illustrate this point we will next construct a condition for a state volume and a corresponding bias potential.

The requirements for defining a state volume are that it contains a region where the probability densities are concentrated and that it must be distinguishable from other such volumes, insofar that it must not share any position vector with these other states. It is desirable that the system exits the state volume only when it makes a transition in order to maximize the boost factor as shown in Eq. (50), but this property is not required and does not affect the accuracy of the hyperdynamics simulation.

If a local variable satisfying the above requirements can be found, then we can construct a bias potential as a function of this local variable. For example, as in Voter's original bias potential, we can use the lowest eigenvalue $\varepsilon_1$ of the Hessian because each region satisfying $\varepsilon_1 > 0$ in the configuration space satisfies these requirements. However, even the most efficient proposed schemes for the eigenvalue calculation require multiple computations of a force vector, and are therefore very computationally intensive. [10, 22]

One possible alternative choice is the potential energy slope or curvature along the direction vector connecting a configuration $\vec{r}$ and the potential energy minimum $\vec{r}_o$. They are defined by

$$s = \nabla V \cdot \vec{u} = \frac{dV}{du} \quad , \tag{55}$$

$$c = \frac{d^2 V}{du^2} \quad , \tag{56}$$

where

$$\vec{u} = (\vec{r} - \vec{r}_o)/l \, , \tag{57}$$

$$l = |\, \vec{r} - \vec{r}_o \,| = \sqrt{\sum_{k=1}^{3N} (r_k - r_{o,k})^2} \quad , \tag{58}$$

where $r_k$ is a component of the position vector $\vec{r}$ and $r_{o,k}$ is a component of the position vector $\vec{r}_o$ at the minimum. In some specific systems, the slope in Eq. (55) and/or the curvature in Eq. (56) become smaller when the system approaches a low probability density region (near the boundary of a state volume) and in such cases it may be possible to determine critical values for them to define the boundary of the state volume. The derivatives are given by

$$\frac{\partial s}{\partial \vec{r}} = \frac{\vec{g}_1}{l} + \left[ \frac{d^2 V}{du^2} - \frac{\vec{g}_1 \cdot \vec{u}}{l} \right] \vec{u} \quad , \tag{59}$$

$$\frac{\partial c}{\partial \vec{r}} = \frac{\vec{g}_2}{l} + \left[ \frac{d^3 V}{d u^3} - \frac{\vec{g}_2 \cdot \vec{u}}{l} \right] \vec{u} \ , \tag{60}$$

where

$$\vec{g}_1 = -\frac{d(\vec{F} l)}{d u} = -\frac{d \vec{F}}{d u} l - \vec{F} \ , \tag{61}$$

$$\vec{g}_2 = -\frac{d^2(\vec{F} l)}{d u^2} = -\frac{d^2 \vec{F}}{d u^2} l - 2 \frac{d \vec{F}}{d u} \ , \tag{62}$$

and

$$\vec{F} = -\nabla V \ . \tag{63}$$

Note that all the higher order derivatives along a specific direction can easily be approximated using a finite difference scheme.

A more inexpensive choice is the hyper-distance from the minimum defined in Eq. (58). The derivative is given by

$$\frac{\partial l}{\partial \vec{r}} = \frac{\vec{r} - \vec{r}_o}{l} \ . \tag{64}$$

Note that the extra computations for this derivative are negligible. If we observe the hyper-distances $l_A$ and $l_B$ from two minima A and B respectively and the probability distributions of $l_A$ and $l_B$ are distinguishable depending on whether the system stays at A or B, then this hyper-distance can be used to construct a bias potential. However, we should approach this method with caution because in some cases the probability distribution of the hyper-distance in one state may not be distinguishable from that in another state. In general the thermal fluctuation of the hyper-distance from a minimum will rise as the number of atoms in the system increases and in some systems transitions occur in a localized region including a relatively small number of atoms. [23] In such cases the hyperdistance between two minima depends only on these few atoms and the hyperdistance associated with the transition is likely to be smaller than the magnitude of the combined thermal fluctuations in the system. However, even in this case the hyper-distance can be used if we can identify where a transition will occur. For example, in the frictional sliding system [23] with a sharp tip scanning a substrate, important transitions always occur at the interface and we can use the atoms at the interface to measure the hyper-distance rather than including all atoms.

## IV. APPLICATION

### A. Model

The methodology described above has been tested with a 3-dimenstional model of an Atomic Force Microscope with a tip and a substrate as shown in Fig. 2 (a). The substrate is a FCC crystal consisting of 1800 atoms and the tip consisting of 183 atoms is made by carving a FCC crystal into a conical shape with flat ends. The tip has the same lattice parameter as the substrate and is joined to the substrate in the [001] direction. As shown in Fig. 2 (b) nine atoms on the bottom layer of the tip are in contact with the substrate. Since the tip and the substrate have the same lattice parameter and are aligned along the same orientation, the tip atoms are in registry with the substrate. All the quantities are expressed in the length units of $\sigma$, the energy units of $\varepsilon$, and the mass units of $m$. Time is measured in units of $\tau = \sigma\,(m/\varepsilon)^{1/2}$. Hereafter the units are omitted unless an ambiguity would arise.

The atoms on the bottom layer of the substrate are fixed to prevent rigid body translation, but they interact with other atoms. Periodic boundary conditions are applied in the horizontal directions. The relative motions of the atoms on the top layer of the tip are constrained, but they can move like a rigid body. The stiffness of an AFM cantilever is modeled by linking a spring with a stiffness $k_S$ of 10 to the top layer of the tip and the spring is elongated by changing the position $x_S$ of a slider located at the end of the spring. A normal force $F_N$ of 5 is also applied downward as shown in Fig. 2 (a).

The interactions of the atoms are modeled by the Lennard-Jones potential,

$$V(r) = 4\,\varepsilon_{ab}\left[\left(\frac{\sigma_{ab}}{r}\right)^{12} - \left(\frac{\sigma_{ab}}{r}\right)^{6}\right], \qquad (65)$$

where $\varepsilon_{ab}$ is the bond energy between the atom of the type $a$ and the atom of the type $b$, $\sigma_{ab}$ is the characteristic length parameter, and $r$ is the distance between the two atoms. We used the following length scale and energy parameters.

$$\sigma_{ss} = \sigma_{tt} = \sigma_{ts} = 1.0\,, \varepsilon_{ss} = \varepsilon_{tt} = 1.0\,,\ \varepsilon_{ts} = 0.2 \qquad (s\text{: substrate, } t\text{: tip})$$

Note that we used a smaller bond energy for the interaction between the tip and the substrate to guarantee that the rearrangement of atom positions always occurs at the interface rather than inside the tip. We observe the dynamics of this system while it is maintained at a constant temperature ($T = 0.01\ \varepsilon/k_B$) by a Nosé-Hoover chain thermostat. [18] The equations of motion are solved using a modified velocity-Verlet algorithm. [25]

In a sliding simulation modeling an AFM experiment [24] the slider moves at a constant velocity, but in this study we used a fixed slider to test the methodology with a non-driven system with constant boundary conditions. The extension of the current methodology to a driven system can be found in Ref.

24. As shown in Fig. 2 (b) the tip can hop to one of eight neighboring positions, each of which defines a metastable state. We biased the relative transition probabilities to these neighboring states by elongating the spring in the positive x direction ($x_S$ = 1.3) so that the hop in the positive x direction dominates other hops. Hereafter the initial state is referred to state A and the final state in the positive x direction is referred to B. We observed a transition from A to B by performing MD simulations with the original potential (conventional simulations) and the biased potentials (hyperdynamics simulations). A transition from state A to state B is detected by performing periodic minimizations using a scheme called FIRE.[26]

From the pre-simulation we have observed that whether the system is at state A or state B can be distinguished by measuring the hyper-distances from each minimum. Therefore, we constructed the bias potentials using the hyper-distance and the following functional form

$$\Delta V(l) = \begin{cases} \Delta V_{\max} & l \le l_L \\ \Delta V_{\max}\left[1-\left(\dfrac{l-l_L}{l_U-l_L}\right)^p\right]^2 & l_L < l < l_U \\ 0 & l \ge l_U \end{cases} , \tag{66}$$

where $l_L = 0.2$, $l_U = 1.0$, and $p = 4$. We prepared four different bias potentials ($V_b^{0.02}$, $V_b^{0.04}$, $V_b^{0.06}$, $V_b^{0.08}$), each of which has $\Delta V_{\max} = 0.02, 0.04, 0.06, 0.08$ respectively. For the original potential and four biased potentials, we simulated 100 samples with different initial conditions.

### B. Results

First, we discuss the results of boost factor calculation. The metastable state volume A is defined as a hyper-sphere in the configuration space, centered at the initial minimum and with a radius of $l_U$. If the probability density function $\rho_W(l)$ of the hyper-distance $l$ in a biased potential $W(= V + \Delta W(l))$ is known, then the biased boost factor of any other biased potential $V_b(= V + \Delta V(l))$, defined in Eq. (32) and Eq. (33), is given by

$$\alpha_b = \frac{\int_0^{l_U} \rho_W(l) e^{+\beta\,\Delta W(l)} dl}{\int_0^{l_U} \rho_W(l) e^{-\beta\,(\Delta V(l)-\Delta W(l))} dl} , \tag{67}$$

where $\rho_W(l)$ is normalized such that

$$\int_0^{l_U} \rho_W(l)\, dl = 1 .$$

For a thermodynamic sampling to obtain $\rho_W(l)$ in a biased potential $W$ we performed a MD simulation with $W$. Moreover, by definition, we must not observe any transition during the sampling. Thus, configurations are periodically stored at prescribed time steps and when a transition is detected we resume the simulation staring from the most recently stored configuration after assigning a new set of Boltzmann-distributed velocities.

Fig. 3 shows the probability distributions of the hyper-distance from the minimum A in the original potential obtained by the MD simulations consisting of 1,000,000 and 5,000,000 integration steps respectively. As the number of MD steps increases, the distribution becomes more accurate. However, even with the larger number of steps (5,000,000), the distribution in the transition region $\widetilde{\mathrm{A}}$ ($l > l_U$) is noisy because this region is rarely visited during the sampling with the original potential.

In Fig. 4, the distributions in various biased potentials ($V_b^{0.02}, V_b^{0.04}, V_b^{0.06}, V_b^{0.08}$) are compared with the distribution in the original potential. All the data were obtained by the MD simulations with 5,000,000 steps except for the distribution in $V_b^{0.08}$, which was obtained after 3,000,000 steps. With an aggressive biased potential like $V_b^{0.08}$, we observed many transitions during the sampling. At these times the sampling had to be resumed from the previously stored configuration resulting in a significantly increased simulation time. One thing to note in these graphs is that as $\Delta V_{\max}$ increases, the probability to be in $\widetilde{\mathrm{A}}$ increases. The role of a biased potential is to force the system to move into this region, increasing the probability for a transition. Moreover, eventually with $V_b^{0.08}$ we have two peaks in the distribution. It is desirable to avoid the creation of such an extra peak because if these two peak regions are separated by very low probability regions, each peak becomes a separate state which can impede ergodically sampling the phase space.

Once we have the probability distribution in one of these potentials ($V$ , $V_b^{0.02}$, $V_b^{0.04}$, $V_b^{0.06}$, $V_b^{0.08}$), we can calculate the probability distribution and the boost factor in any other potential. Fig. 5 shows the biased boost factors in $V_b^{0.02}$, $V_b^{0.04}$, $V_b^{0.06}$, $V_b^{0.08}$ as functions of integration steps, and each graph has two curves: one is obtained from the original potential and the other from the corresponding biased potential. For the same reason above, the biased boost factor in $V_b^{0.08}$ calculated from the distribution in $V_b^{0.08}$ was obtained only up to 3,000,000 steps. In principle these two boost factors from the original potential and the biased potential should converge as the number of steps increases, but as shown in Fig. 5 the difference becomes large as $\Delta V_{\max}$ increases. This can be explained by referring to Fig. 6, which shows the probability distributions of the hyper-distance in $V_b^{0.08}$ recovered from the original potential and the biased potentials. If we obtained the exact distribution in each sampling with a specific potential, the recovered distributions in other potentials should also be exact. However, because we use a finite number of steps for a sampling, there is always an inaccuracy in the distribution. This inaccuracy also makes the error in the boost factor calculation larger as $\Delta V_{\max}$ increases because of the exponential term in boost factor calculation.

As an alternative sampling, we used the umbrella sampling method. In the umbrella sampling instead of using either the original potential or one of the biased potentials for the hyperdynamics simulations, we used harmonic potentials given by

$$\begin{aligned} V_h^i &= V + \Delta V_h^i \\ \Delta V_h^i &= \frac{1}{2} k^i (l - l^i)^2 \qquad (i = 1, 2, 3, 4, 5) \end{aligned} , \tag{68}$$

where $k^1 = k^2 = k^3 = k^4 = k^5 = 1.0$, and $l^1 = 0.4$, $l^2 = 0.6$, $l^3 = 0.8$, $l^4 = 1.0$, $l^5 = 1.2$. Each harmonic potential is designed to sample the region near its center more accurately. For each harmonic potential we sampled 1,000,000 points and once we have the probability distribution of the hyper-distance in each harmonic potential, we can reconstruct the distribution in the original potential using WHAM as shown in Fig. 7. [20] Although we used the same number of MD steps both in the conventional sampling and in the umbrella sampling ($5 \times 1{,}000{,}000 = 5{,}000{,}000$), we obtained a more accurate result with the umbrella sampling.

Fig. 8 shows the biased boost factors of the hyperdynamics biased potentials ($V_b^{0.02}$, $V_b^{0.04}$, $V_b^{0.06}$, $V_b^{0.08}$). The boost factors calculated from the conventional sampling with the original potential and the hyperdynamics biased potentials are compared with the boost factor obtained from the umbrella sampling. The results with $V_b^{0.08}$ significantly deviate from the umbrella sampling boost factors because of the reasons explained above.

Next, we performed the MD simulations to see the transition from A to B with the original potential and the biased potentials. For each potential we prepared 100 samples with different initial conditions and measured the waiting times. Fig. 9 (a) and (b) show the average waiting time in the state volume ($l < l_U$), $\overline{t_{A,i}}$, and the average waiting time in the transition region ($l > l_U$), $\overline{t_{A,o}}$. As $\Delta V_{\max}$ increases, that is, the boost factor increases, $\overline{t_{A,i}}$ decreases, but $\overline{t_{A,o}}$ remains almost within the error bar, which is the standard error, regardless of the potentials as we discussed in Sec. II. C. Fig. 10 shows the average number of crossings of $\partial A(l = l_U)$, the boundary of the state volume, before a transition. It also remains within the standard error, which confirms $f_{A\to B} = (f_{A\to B})_b$.

Finally, Fig. 11 shows the original waiting times recovered from the waiting times measured in the hyperdynamics simulations with the biased potentials through two different methods. In one method we use the boost factor obtained from the pre-simulation as discussed in Sec. III. A, the original waiting time is recovered using the relation, $\alpha_b \times (t_{A,i})_b + (t_{A,o})_b$, and the other method by Voter uses Eq. (44), that is, the boost factor is calculated during the simulation on the fly. The original waiting time is well recovered from the waiting times in $V_b^{0.02}$, $V_b^{0.04}$, $V_b^{0.06}$, but the recovered waiting time from the simulations with $V_b^{0.08}$ deviates with a larger error even with the pre-calculated boost factor. We conjecture that it is because the biased waiting time in $V_b^{0.08}$ becomes too short to satisfy the criterion that the system is fully equilibrated in a state before making a transition. The waiting times recovered using Eq. (44) show larger deviations from the results obtained from the pre-calculated boost factor.

## V. CONCLUSION

We have reformulated the hyperdynamics method, which was devised to accelerate the conventional MD simulation, in a rigorous way that does not require a TST dividing surface. First, we define a transition as an event between two state volumes in the configuration space rather than a crossing of a TST dividing surface. By the fact that the ratio of the number of transitions to the number of crossings of the boundary of these state volumes remains unchanged in a biased potential, we showed that the boost factor, which is the ratio of the transition rates in the biased potential and in the original potential, can be used to exactly recover the original waiting time.

We presented a new perspective to see the boost factor as a multiplication factor between the waiting times in the biased potential and in the original potential so that it can be calculated in a pre-simulation rather than during the simulation as in Voter's original method. To accurately calculate the boost factor we discussed the various aspects of thermodynamic sampling and concluded that the umbrella sampling gives a more accurate result when compared with the conventional sampling performed "on-the-fly".

Moreover, with the criteria of a state volume rather than using a TST dividing surface, we devised new bias potentials using local variables other than the eigenvalue and the eigenvector of the Hessian matrix. Among these local variables the hyper-distance from a minimum turned out to be the most efficient. However, an important *caveat* must be noted. The hyper-distance can only be used in a system where the hyper-distances between minima are distinguishable from the thermal fluctuation of the hyper-distance from each minimum, a condition that holds for our AFM simulations because a confined transition region can be identified *a priori*. The new methodology was tested for an AFM system modeled by the Lennard-Jones potential, and the simulation gave the results consistent with the theory.

## ACKNOWLEDGEMENT

WKK and MLF acknowledge support of the NSF program on Materials and Surface Engineering under Grants CMMI-0510163 and CMMI-0926111 and the use of facilities at the Johns Hopkins University Homewood High Performance Compute Cluster.

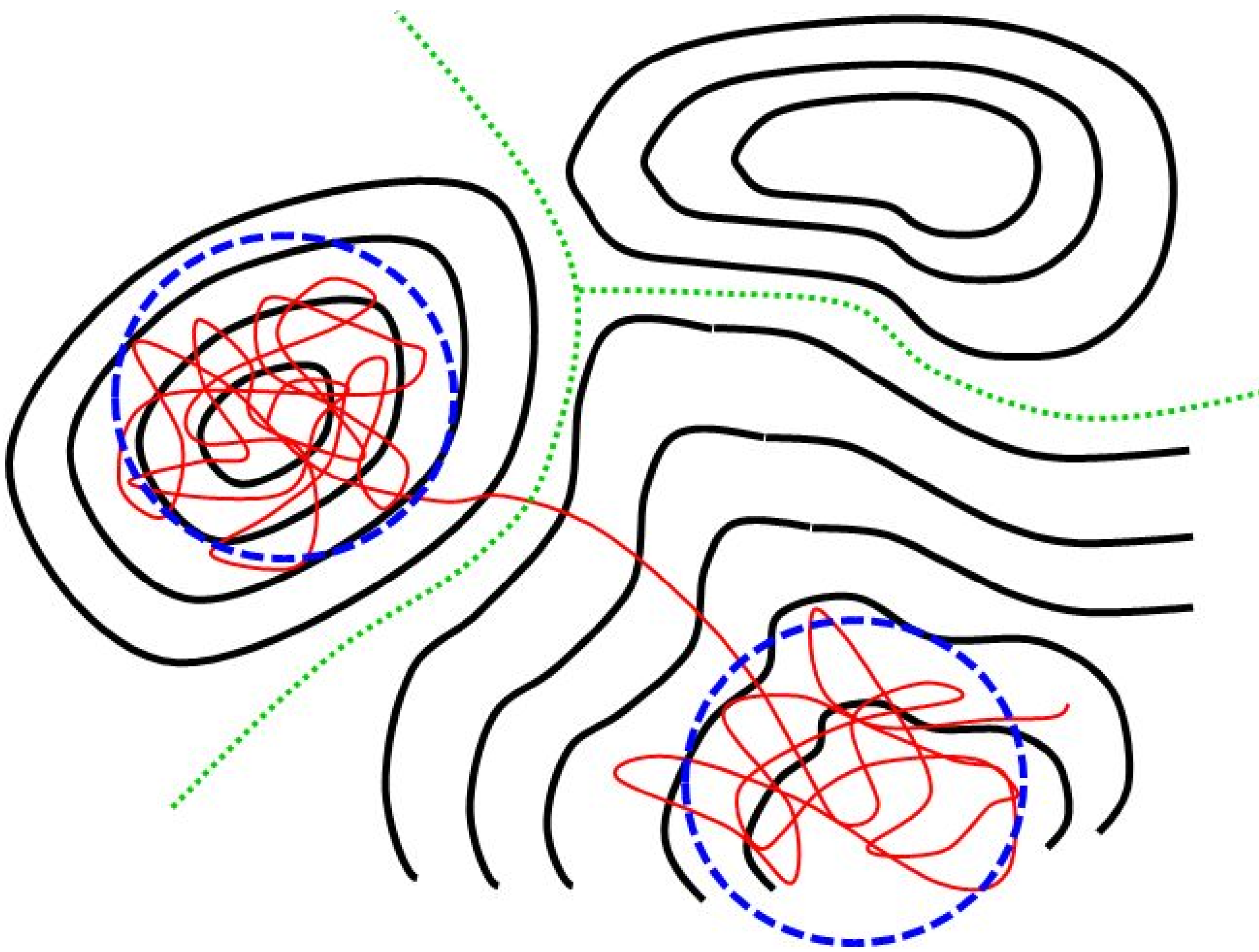

FIG. 1. An example of 2-dimensional potential energy contours (thick solid curves). The entangled light solid curve represents a system trajectory, the dashed circles represent the boundaries of state volumes, and the dotted curve is a TST dividing surface. Note that a transition from one metastable state to another is evident although several crossings of the boundaries of these state volumes occur before the system makes a transition.

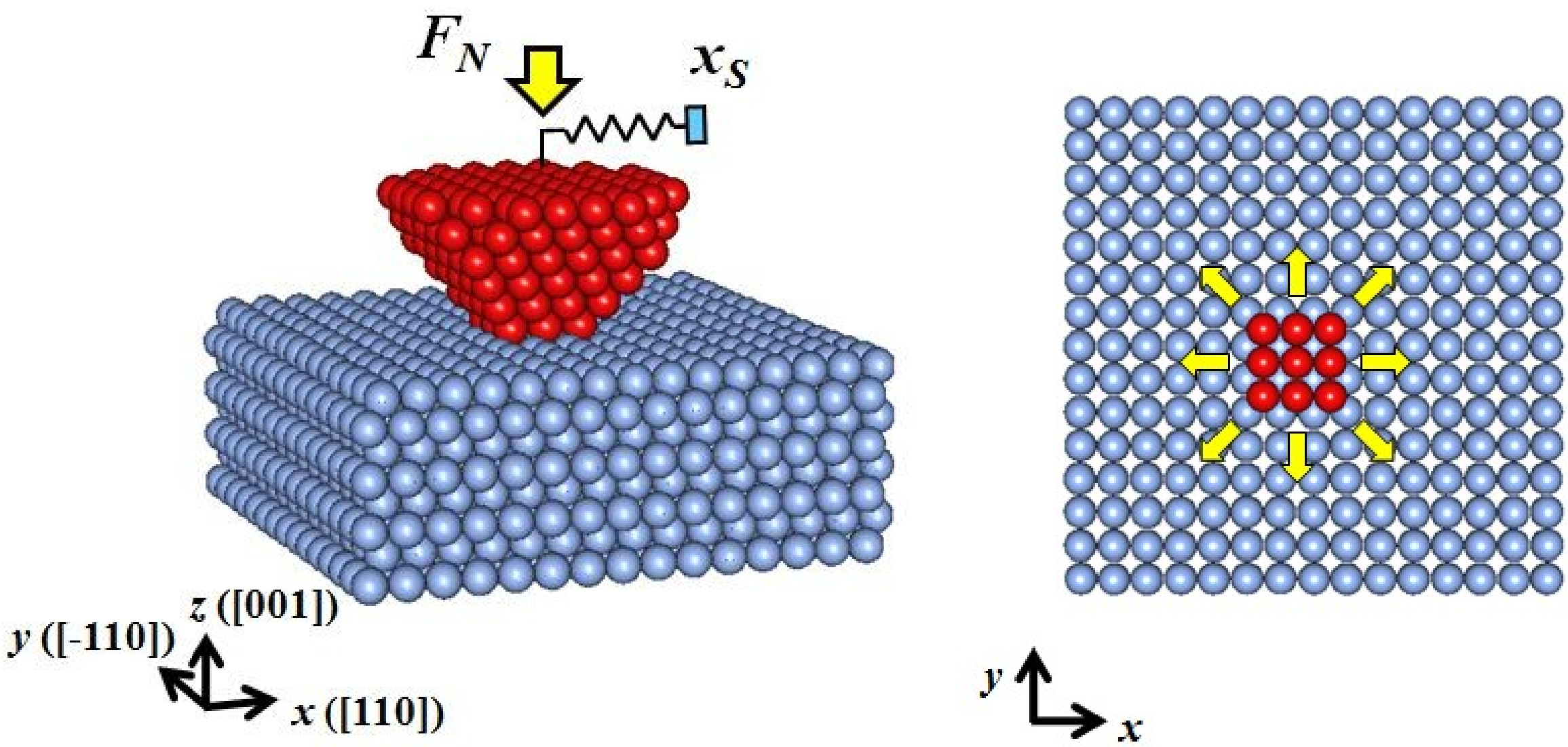

(a) (b)

FIG. 2. A diagram of 3-dimensional AFM model consisting of a tip and a substrate. (a) The normal force and the spring force are applied on the top layer of the tip and the spring force can be controlled by changing the slider position $x_s$. (b) The configuration of atoms at the interface between the bottom layer of the tip and the top layer of the substrate. The arrow shows the direction for a transition to one of the neighboring states.

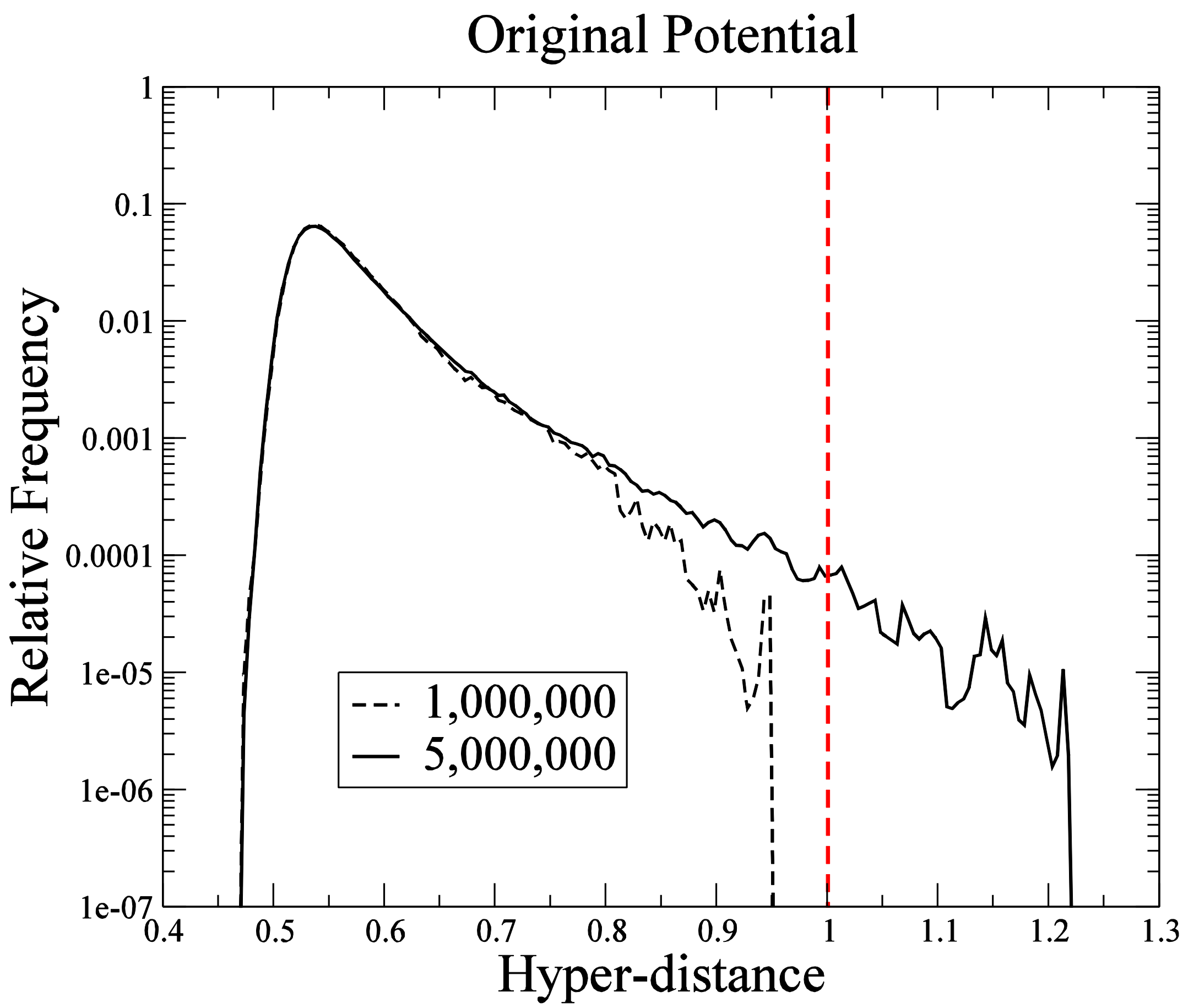

FIG. 3. The probability distribution of the hyper-distance from the minimum A in the original potential. The dashed curve was obtained after 1,000,000 MD steps and the solid curve was obtained after 5,000,000 MD steps. The dashed vertical line shows the boundary of the state volume A.

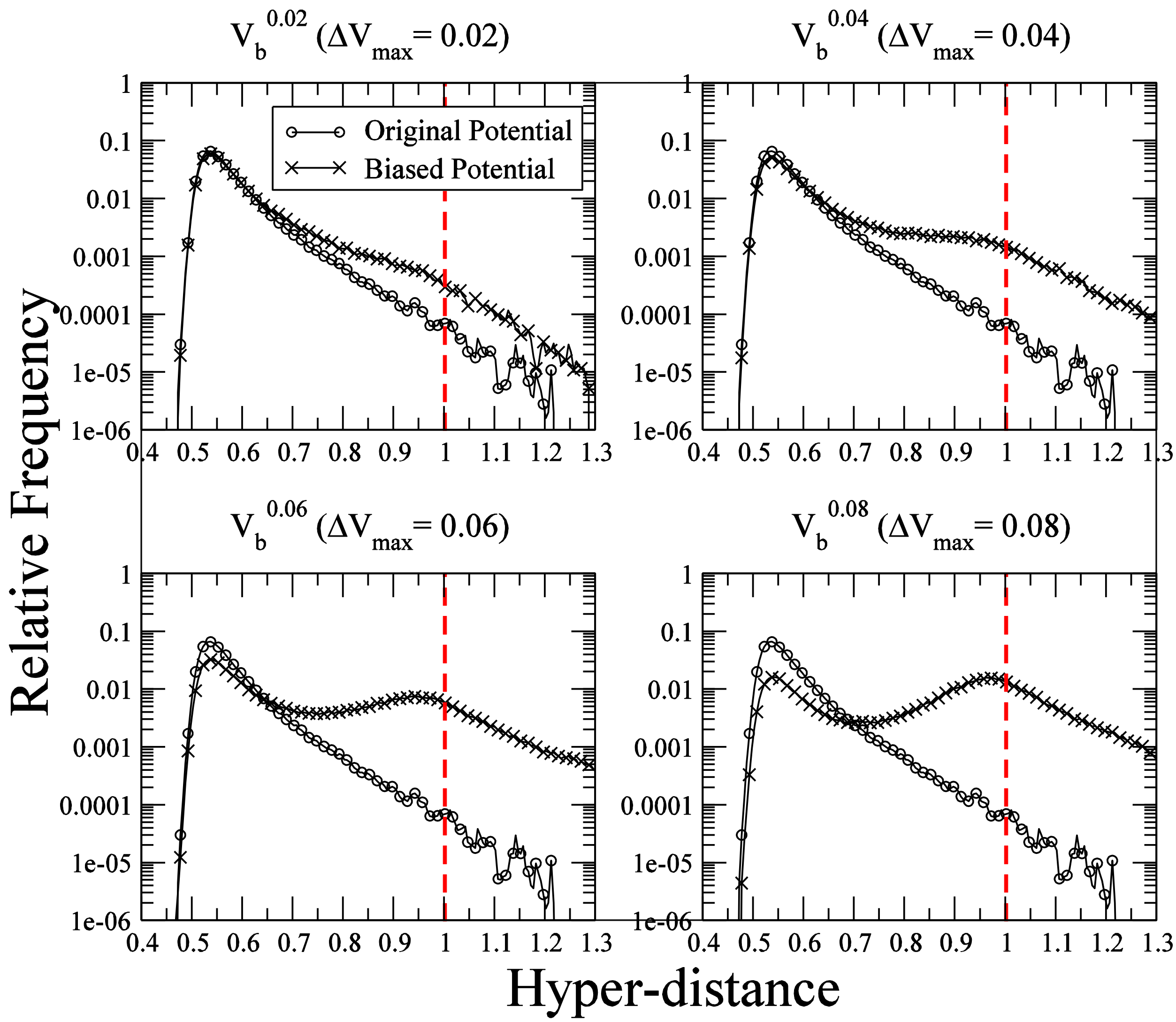

FIG. 4. The probability distributions of the hyper-distance from the minimum A in the biased potentials ($V_b^{0.02}, V_b^{0.04}, V_b^{0.06}, V_b^{0.08}$). For comparison, the distribution in the original potential is plotted together with the distribution in the biased potentials. All the results were obtained by the MD simulations with 5,000,000 steps except for the distribution in $V_b^{0.08}$, which was obtained after 3,000,000 steps.

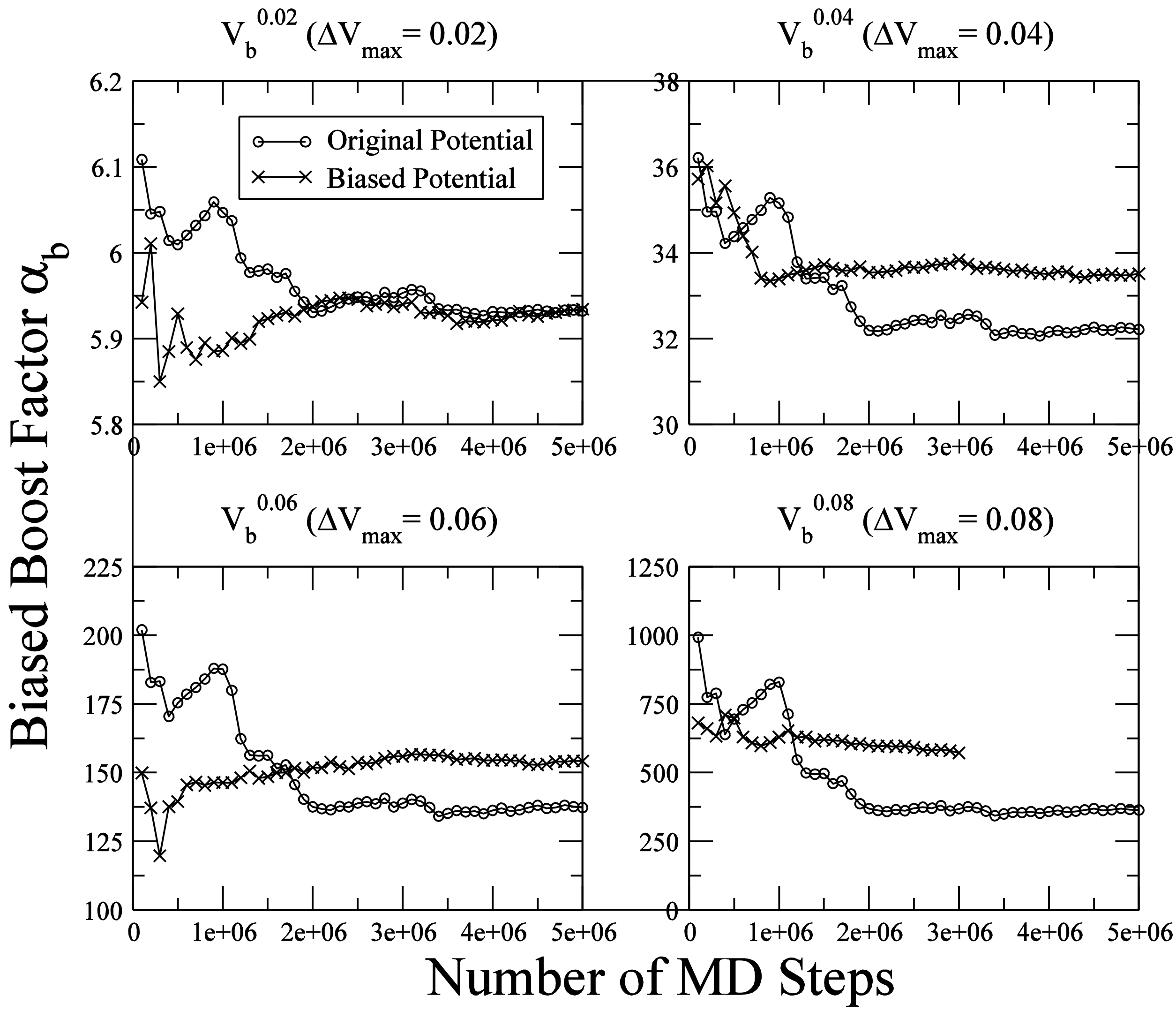

FIG. 5. The biased boost factors in the biased potentials ($V_b^{0.02}, V_b^{0.04}, V_b^{0.06}, V_b^{0.08}$). For comparison, the biased boost factor calculated from the hyper-distance distribution in the original potential is plotted together with the biased boost factor obtained from the hyper-distance distributions in the corresponding biased potentials.

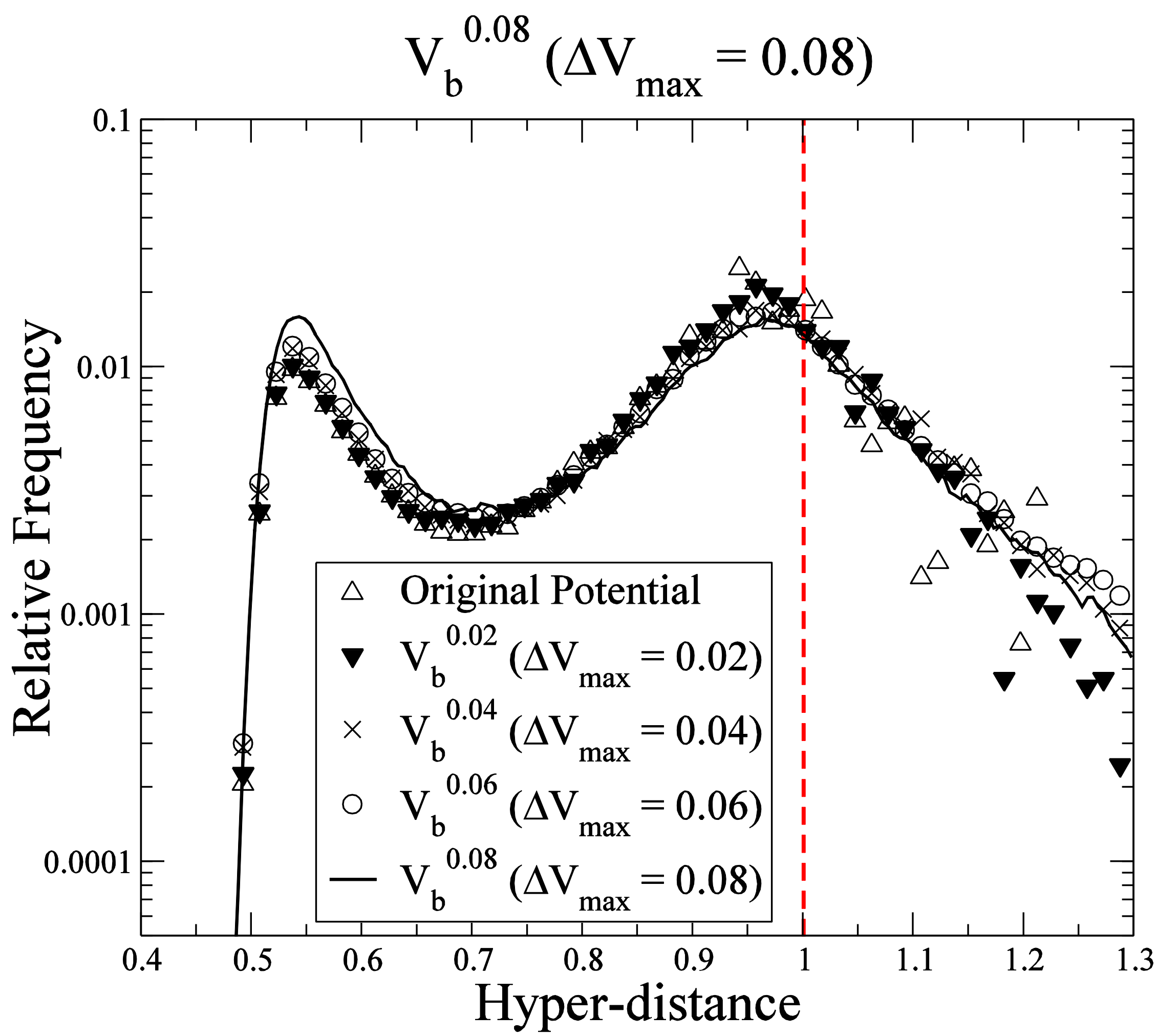

FIG. 6. The probability distributions of the hyper-distance in $V_b^{0.08}$. One was obtained from the simulation with $V_b^{0.08}$ and the others were recovered from the distributions in the original potential and $V_b^{0.02}, V_b^{0.04}, V_b^{0.06}$.

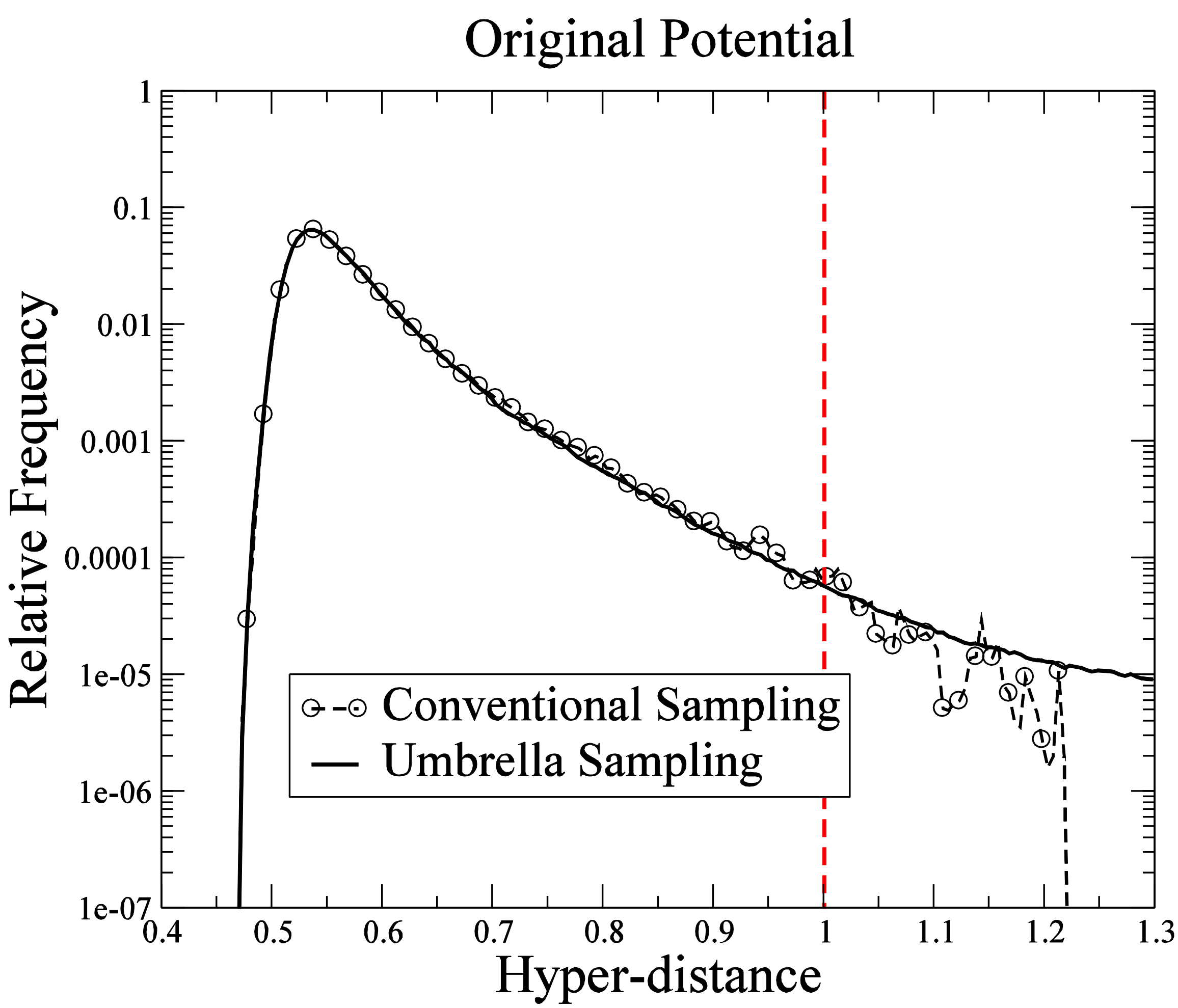

FIG. 7. The probability distribution of the hyper-distance in the original potential obtained from the conventional sampling and the umbrella sampling using 5 harmonic potentials.

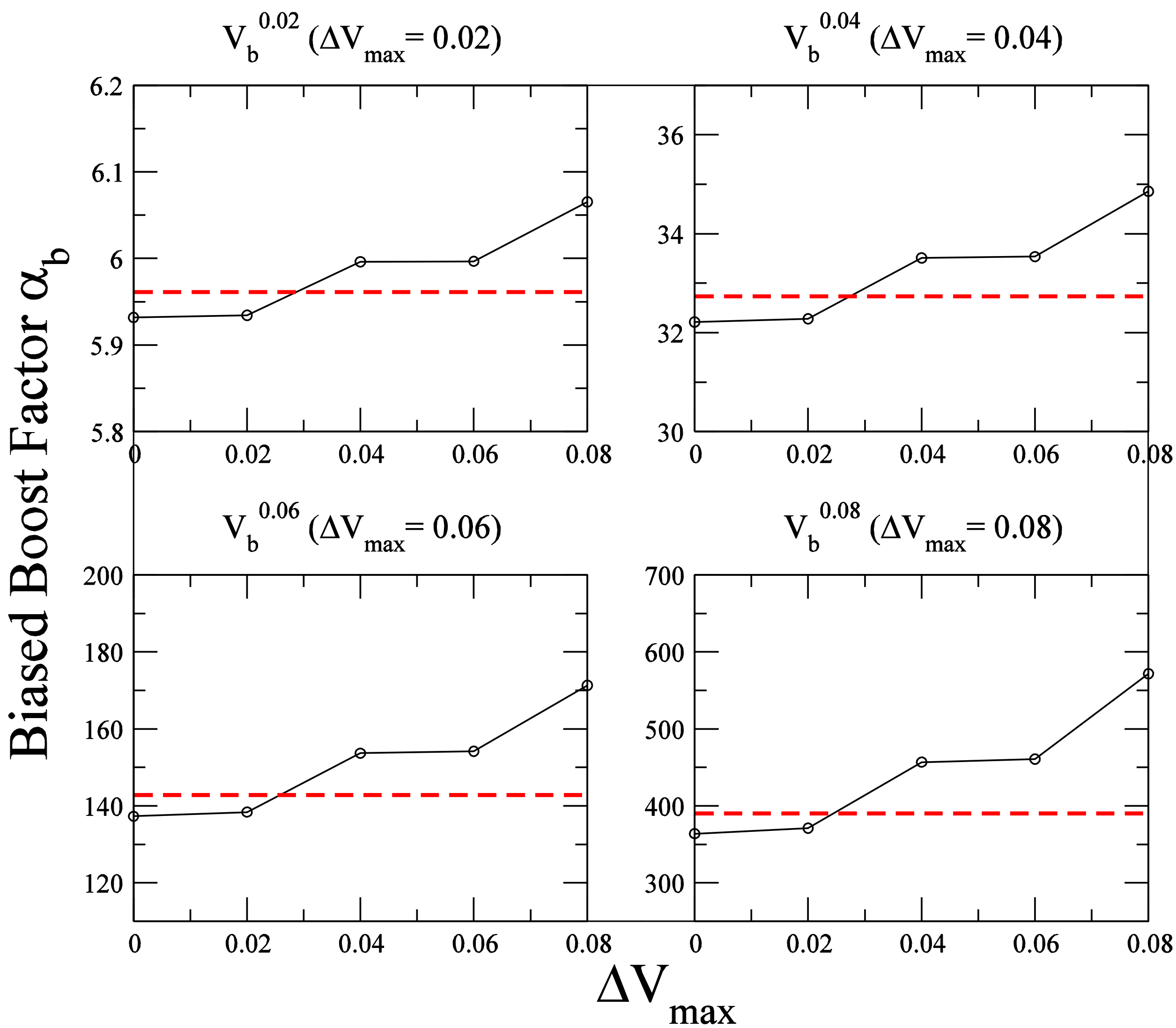

FIG. 8. The biased boost factors of the biased potentials ($V_b^{0.02}, V_b^{0.04}, V_b^{0.06}, V_b^{0.08}$). The open dots show the boost factors obtained from the conventional sampling with the potential whose $\Delta V_{max}$ is given in the x axis and the solid dashed lines show the boost factors obtained from the umbrella sampling.

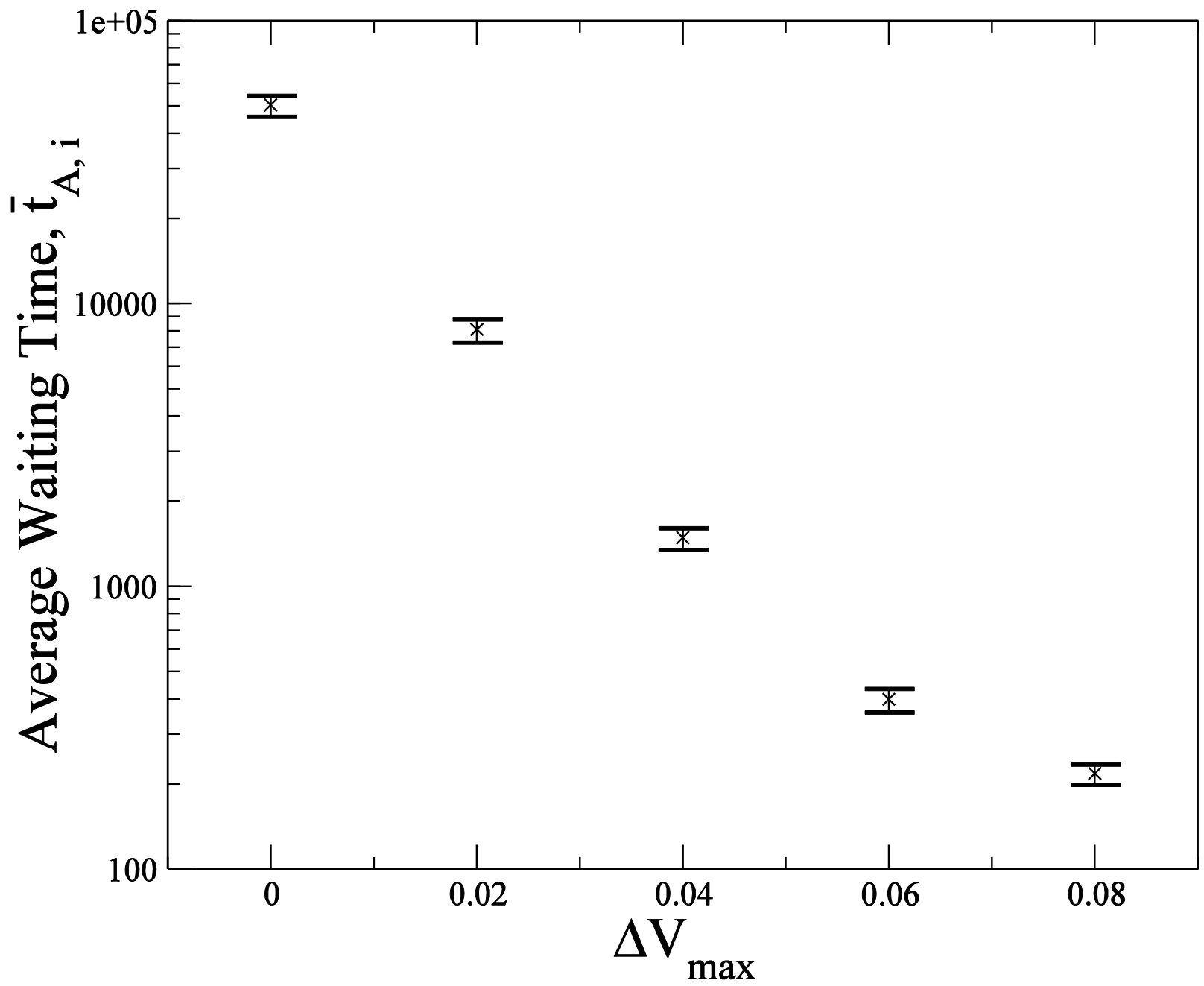

(a)

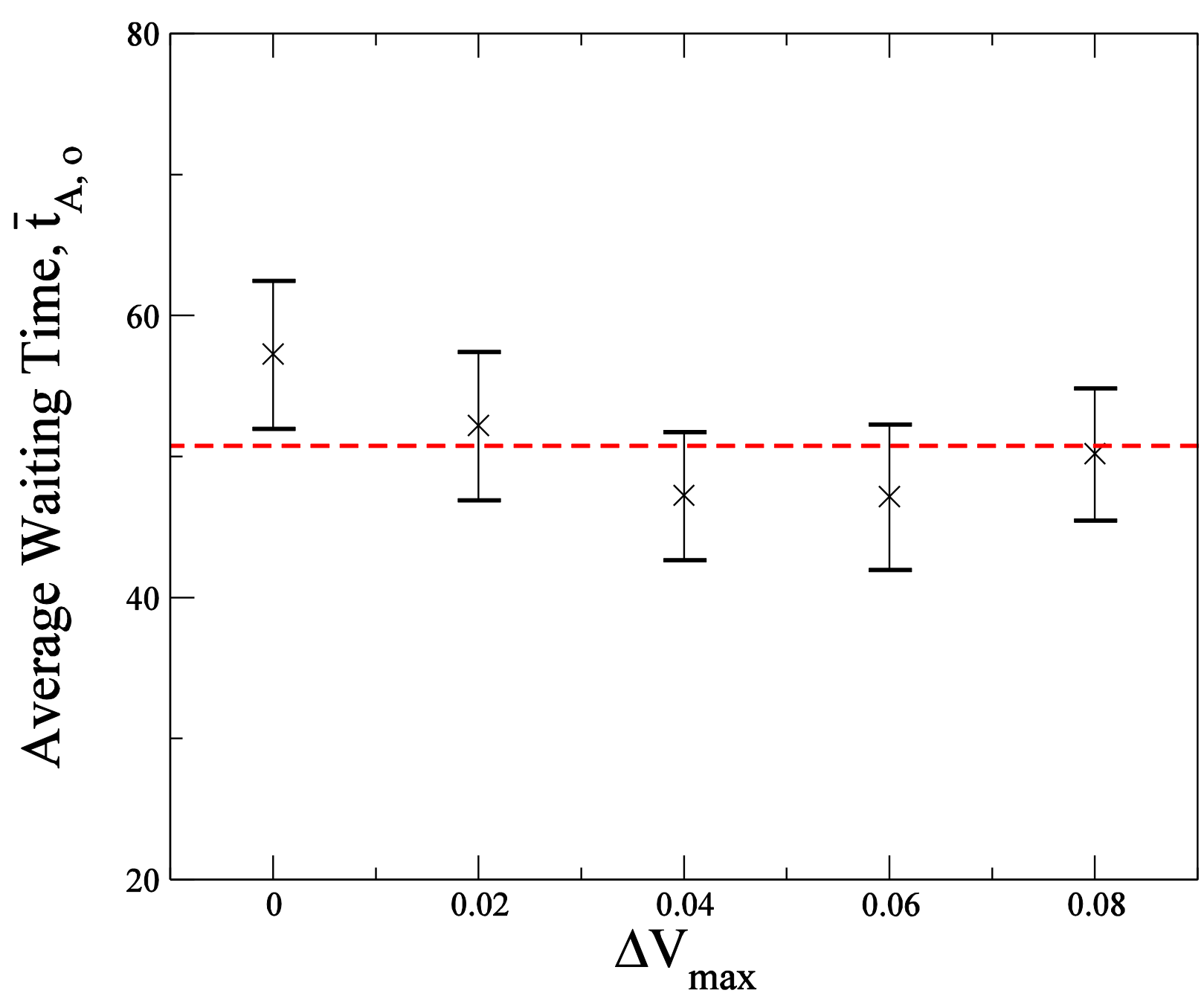

(b)

FIG. 9. The average waiting times in the original potential and the biased potential whose $\Delta V_{\max}$ is given in the x axis. (a) The average waiting time in the state volume A, $\overline{t_{A,i}}$ . (b) The average waiting time in the transition region $\tilde{A}$ , $\overline{t_{A,o}}$ . The red dashed line is the total average and the error bars show the standard error.

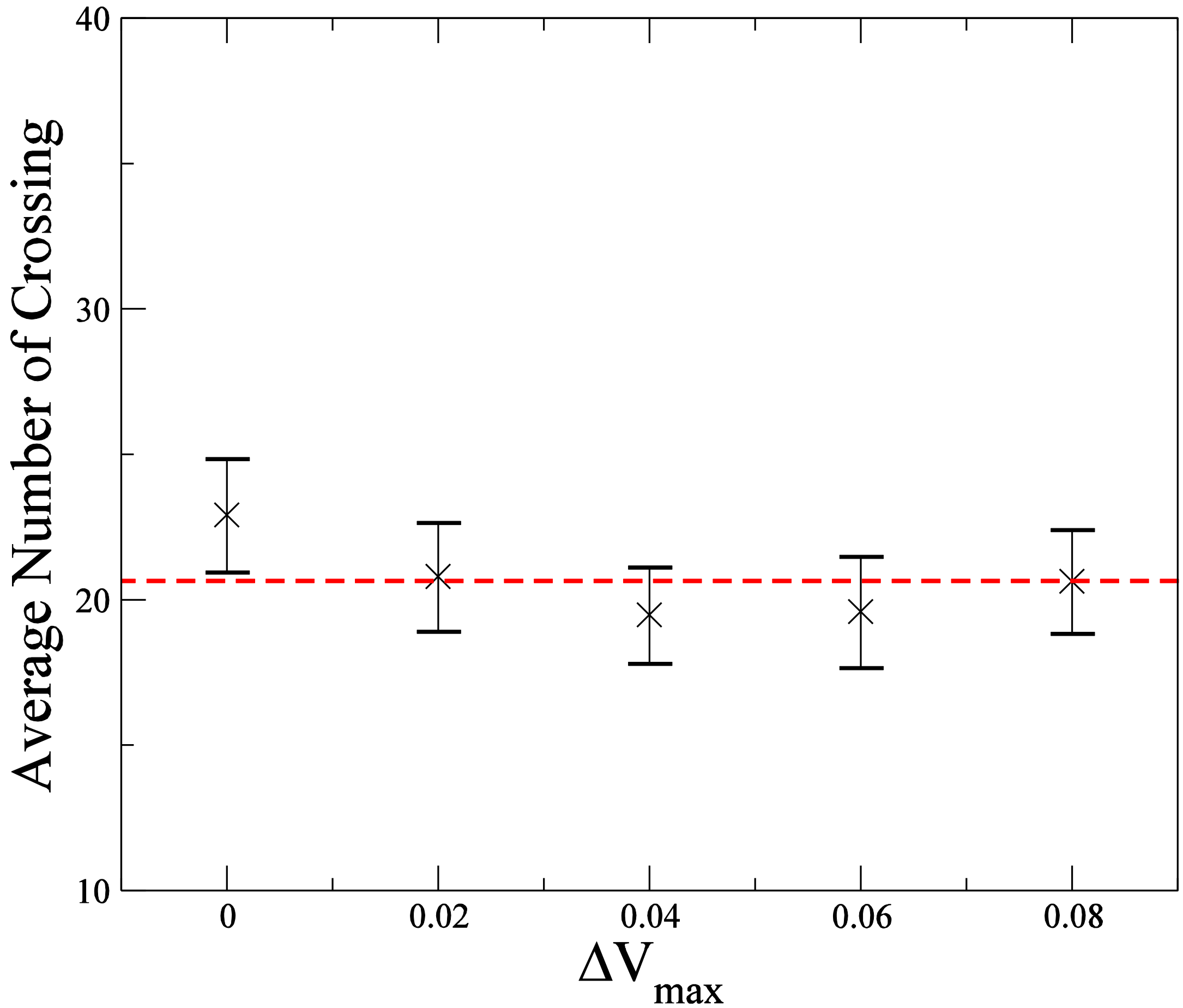

FIG. 10. The average number of crossing of the boundary of the state volume A ($\partial$A) in the original potential and the biased potential whose $\Delta V_{\max}$ is given in the x axis. The red dashed line is the total average and the error bars show the standard error.

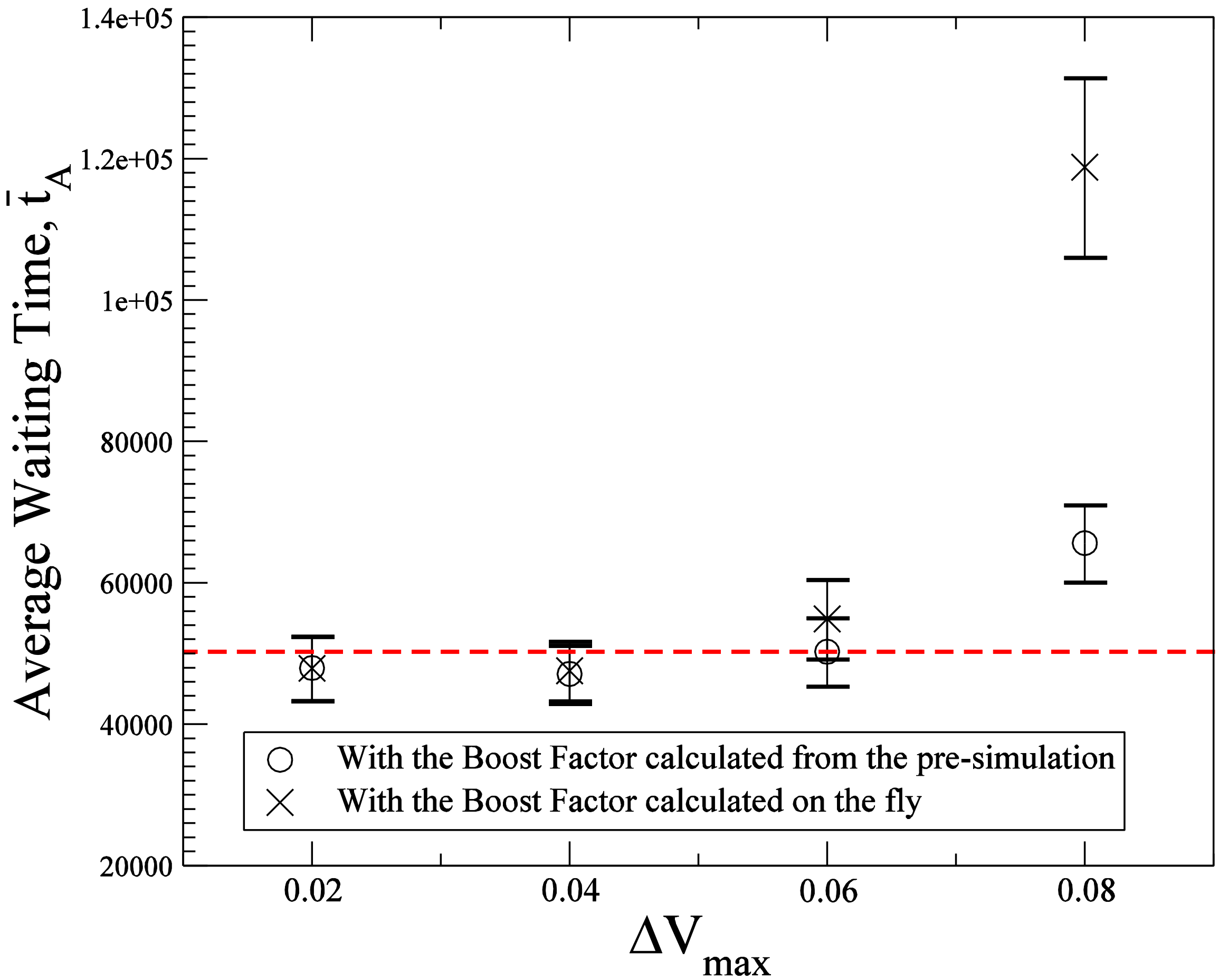

FIG. 11. The average of the original waiting times recovered from the waiting times in the biased potentials, whose $\Delta V_{max}$ is given in the x axis, with the boost factors calculated via two different schemes; one is the pre-simulation and the other is the on-the-fly calculation. The red dashed line is the average waiting time in the original potential and the error bars show the standard error.